# Engineering symmetry breaking in two-dimensional layered materials

Luojun Du[1]*, Tawfique Hasan[2], Andres Castellanos-Gomez[3], Gui-Bin Liu[4], Yugui Yao[4], Chun Ning Lau[5], Zhipei Sun[1,6]*

[1]Department of Electronics and Nanoengineering, Aalto University, Tietotie 3, FI-02150, Finland

[2]Cambridge Graphene Centre, University of Cambridge, Cambridge, CB3 0FA, UK

[3]Materials Science Factory, Instituto de Ciencia de Materiales de Madrid (ICMM-CSIC), Madrid, E-28049 Spain

[4]Key Lab of Advanced Optoelectronic Quantum Architecture and Measurement (MOE), School of Physics, Beijing Institute of Technology, Beijing 100081, China

[5]Department of Physics, The Ohio State University, Columbus, OH 43210, USA

[6]QTF Centre of Excellence, Department of Applied Physics, Aalto University, FI-00076 Aalto, Finland

*Corresponding authors: luojun.du@aalto.fi; zhipei.sun@aalto.fi

**Symmetry breaking in two-dimensional layered materials plays a significant role in their macroscopic electrical, optical, magnetic and topological properties, including but not limited to spin-polarization effects, valley-contrasting physics, nonlinear Hall effects, nematic order, ferroelectricity, Bose-Einstein condensation and unconventional superconductivity. Engineering symmetry breaking of two-dimensional layered materials not only offers extraordinary opportunities to tune their physical properties, but also provides unprecedented possibilities to introduce completely new physics and technological innovations in electronics, photonics and optoelectronics. Indeed, over the past 15 years, a wide variety of physical, structural and chemical approaches have been developed to engineer symmetry breaking of two-dimensional layered materials. In this Technical Review, we focus on the recent progresses on engineering the breaking of inversion, rotational, time reversal and spontaneous gauge symmetries in two-dimensional layered materials, and illustrate our perspectives on how these may lead to potential new physics and applications.**

## 1. Introduction

Since the first exfoliation of graphene, the unique properties of two-dimensional (2D) van der

Waals (vdW) materials that are unattainable with other materials have initiated the '2D Age' and transformed the landscape of fundamental research and technological advances in physics, chemistry, energy harvesting, information and beyond[1-5]. A rich variety of exotic physical phenomena have thus far been demonstrated in 2D materials, including but not limited to nonlinear optics[6-10], Hall effects[11-14], valley-contrasting physics[4,15-17], spin-orbit physics[18-20], chiral quasiparticles[21-24], piezoelectricity[25-27], superfluidity and superconductivity[28-31]. Remarkably, majority of these fascinating physical phenomena are fundamentally dictated by the broken symmetry in 2D materials, e.g., the breaking of inversion, rotational, time-reversal and gauge symmetries (Box 1). For example, circular photogalvanic effect[32,33] (nonreciprocal $c$-type second harmonic generation[6,34]), light-valley interactions[4] (magneto-circular dichroism[35,36]), nonlinear Hall effect[11,14] (quantum anomalous Hall effect[12,37]), ferroelectricity[38,39] (topological spin excitations[40-42]) and unconventional Ising superconductivity[29,31,43] (valley Zeeman effect[44-47]) are closely related to the broken inversion (time-reversal) symmetry. Consequently, it is of particular importance to engineer the symmetry breaking of 2D materials, which could enable the control of these fascinating physical properties and thus set a foundation for the development of various emerging technological advances. In addition, controlled manipulation of symmetry breaking offers exciting possibilities to manipulate the internal quantum degrees of freedom (e.g., spin, valley and layer pseudospin) in 2D materials and therefore provide a firm basis for exotic spintronics[48], valleytronics[4,49] and twistronics[50]. Further, manipulation of symmetry breaking in 2D materials can create unique opportunities to integrate different broken symmetries into the same system, providing an unprecedented path to underpin nearly limitless new physics and possibly initiate new eras in technological innovations.

Over the past decade, substantial developments and progresses on engineering the symmetry breaking of atomically thin 2D materials have been witnessed. Remarkably, 2D materials can be integrated with other 2D/non-2D materials to create all-2D/mixed-dimensional vdW heterostructures via weak non-covalent interactions without the constraints in traditional heterostructures (e.g, lattice-mismatching, interfacial defects and interlayer atomic diffusion)[2,51-53]. This therefore offers a distinctive way to engineer symmetry breaking in 2D materials through interlayer twist angle and strong layer-layer many-body interactions, and provides a powerful platform to harness the synergies of different dimensionalities. Here, we review the state-of-the-art engineering approaches to break the inversion, rotational, time reversal and spontaneous gauge symmetries in 2D materials. We also conclude with perspectives on the engineering of explicit and spontaneous symmetry breaking in 2D materials and its implications on new physics and novel technological applications.

**Box 1 | Fundamentals of symmetry breaking**

**Inversion symmetry breaking:** Inversion symmetry implies that there is at least one point (e.g., the center of symmetry) which makes the crystal under the operation of inversion symmetry $r(x,y,z) \rightarrow r(-x,-y,-z)$ look the same as the original. If there is no symmetry center in a crystal, it suggests that the inversion symmetry is broken. Out of the 32 point groups, 21 are non-centrosymmetric including 10 polar point groups. The inversion symmetry breaking dictates many enchanting physical phenomena, e.g., even-order nonlinear effects, spin-orbit physics, Weyl fermions, valley-contrasting effects, ferroelectricity and unconventional Ising superconductivity.

**Rotational symmetry breaking:** Rotational symmetry, a spatial symmetry, means that the crystal looks exactly the same after a partial turn around a rotation axis. In contrast to amorphous solids that possess the invariance under arbitrary rotation, crystals break rotational symmetries down to a discrete subgroup. For instance, crystals can harbor only 1-fold ($C_1$), 2-fold ($C_2$), 3-fold ($C_3$), 4-fold ($C_4$), and 6-fold ($C_6$) rotational symmetries. The $C_6$ and $C_3$ rotational symmetries typically ensure the isotropic properties of a matter. When $C_3$ rotational symmetry is broken, the crystal could exhibit intriguing in-plane anisotropies in vibrational, thermal, optical and electrical properties.

**Time reversal symmetry breaking:** Time reversal symmetry describes the immutability under the transformation of time reversal $T: t \rightarrow -t$. As $T$ is an anti-unitary operator, the wave function becomes the complex conjugation under $T$: $T\psi(r,t) = \psi^*(r,-t)$. In other words, the wave vector $k$ of Bloch state changes to $-k$ under the operation of $T$. As a result, the physical properties (such as energy eigenvalue $E$ and Berry curvature $\Omega$) at wave vectors $k$ and $-k$ are linked by time reversal symmetry. Time reversal symmetry breaking can lift the degeneracy and underlies various fascinating physical phenomena, such as exotic spin excitations, quantum anomalous Hall and magneto-optical effects.

**Gauge symmetry breaking:** Compared with the space-time symmetries, gauge symmetry is a subtle kind. Gauge symmetry refers to the phase invariance of wave function under the operation of phase transformation ($\boldsymbol{\psi} \to e^{i\alpha}\boldsymbol{\psi}$). If the phase of wave function changes, it signifies the broken gauge symmetry. The spontaneous gauge symmetry breaking lays a foundation for the exotic superfluidity and superconductivity.

## 2. Engineering of inversion symmetry breaking in 2D materials

Inversion symmetry breaking provides the prerequisite for a large portfolio of fascinating physical phenomena, such as valley-contrasting physics and Rashba/Dresselhaus spin-orbit coupling[4,48]. A wide range of methods have been demonstrated to manipulate and engineer the inversion symmetry breaking in 2D materials. This facilitates the unprecedented possibility of continuously tuning the physical properties (e.g., Berry curvature, valley magnetic moment and electronic bandgap) for various exciting applications[54-57]. In this section, we discuss the most widely-explored approaches to manipulate the inversion symmetry breaking of 2D materials.

### 2.1 Electric field

Electric field provides a powerful tool to break the inversion symmetry of 2D crystals due to its polar nature. For instance, when a vertical electric field is applied onto a Bernal-stacked centrosymmetric bilayer graphene with a dual-gated architecture (a top and bottom gates, left panel of Fig. 1a), it introduces an asymmetric interlayer electrostatic potential between the upper and lower layers, breaking the inversion symmetry (right panel of Fig. 1a)[58-63]. Note that the dual-gated device structure is the most commonly used configuration to apply out-of-plane electric fields in 2D materials while keeping other conditions (such as carrier density) unchanged[55,64]. Inversion symmetry breaking controlled by the perpendicular electric field can open a bandgap between the conduction and valence bands in bilayer graphene (left panel of Fig. 1b)[58,64]. Of particular importance, the degree of inversion symmetry breaking and the opened bandgap (up to ~ 250 meV[55,65]) in bilayer graphene are highly tunable with the magnitude and direction of electric field, showing great interests for applications (e.g., large current on/off ratio of up to $10^6$ for nanoelectronics with ultralow standby power dissipation[66] and high-sensitivity hot-electron bolometers[67]).

Further, inversion symmetry breaking engineered by out-of-plane electric field can produce valley-contrasted Berry curvatures $\boldsymbol{\Omega}(\boldsymbol{k})$ and orbital magnetic moment[57,64,68], leading to fascinating valley-dependent electrical transport[49,64] and optical selection rules[69-71], respectively. In particular, Berry curvature can serve as an out-of-plane internal magnetic field in momentum space and drive an anomalous transverse velocity $\boldsymbol{v_a}$ under an applied electric field $\boldsymbol{E}$: $\boldsymbol{v_a} = \frac{e}{\hbar} \boldsymbol{E} \times \boldsymbol{\Omega}(\boldsymbol{k})$, where $e$ and $\hbar$ denote the elementary charge and the reduced Planck constant, respectively[49,72]. Since Berry curvatures possess opposing sign in opposite valleys, carriers in different valleys gain opposing transverse velocities, giving rise to the pure valley current and valley Hall effect[4,15]. Recently, continuously tunable Berry curvature (right panel of Fig. 1b) and valley Hall effect have been uncovered in bilayer graphene by controlling the inversion symmetry breaking with perpendicular electric fields[57,64]. Moreover, by manipulating the inversion symmetry breaking with vertical electric fields, topologically protected one-dimensional chiral states can occur at the domain walls between AB- and BA-stacked bilayer graphene[73,74] or the line junction between band-inverted bilayer graphene insulators[75,76]. Such chiral edge states of quantum valley Hall insulators are topologically protected against elastic backscattering by time-reversal symmetry, opening opportunities towards the realization of topological valleytronic devices with ultralow dissipation. Apart from bilayer graphene, the perpendicular electric field can also be applied onto a wide variety of centrosymmetric 2D crystal structures (e.g., monolayer $WTe_2$[32,77], bilayer $MoS_2$[54,78] and rhombohedral stacked trilayer graphene[79,80]) to controllably break the inversion symmetry. This permits the realization of a wealth of continuously tunable physical properties (such as electrically switchable Berry curvature dipole[32,77,81], modulated Zeeman-type spin splitting[82] and tunable second-order electric susceptibility[78,83]), attractive for technological advances in electronics, photonics and valleytronics.

**2.2 Staggered sublattice potential**

Staggered sublattice potential induced by a substrate can impose different onsite energies onto different sublattices and thus break the sublattice equivalence and inversion symmetry[49]. For instance, monolayer graphene harbors inversion symmetry due to its honeycomb structure composed of two crystallographically equivalent triangular carbon sublattices (termed as the A and B sublattices). When monolayer graphene is placed on top of a *h*-BN substrate with near zero angle mismatch, the lattice of monolayer graphene would relax and reconstruct to minimize the stacking potential energy[84-86]. The resulting energetically favourable configuration belongs to the commensurate state and possesses one carbon sublattice on top of the boron and other carbon sublattice atop the nitrogen (Fig. 1c)[84,85]. The different nature between boron and nitrogen gives rise to the atomic-scale modulations of onsite electric potential between the A and B carbon sublattices, leading to the inequivalence between the two sublattices and hence breaking the inversion symmetry[85-88]. The broken inversion symmetry in monolayer graphene controlled by the staggered sublattice potential of *h*-BN can lead to strong second-order nonlinear responses[89] and open energy gaps at both the primary Dirac points (PDP) and secondary Dirac cones (SDC) induced by Bragg scattering from the long-wavelength moiré superlattice potential[50,56,86,90]. However, it should be pointed out that the bandgaps in graphene/*h*-BN with near 0° alignment (~40/25 meV at the PDP/SDC) obtained by transport[50] are much smaller than that revealed by angle-resolved photoemission spectroscopy[90]. Such disagreement regarding the gap size may be attributed to charge inhomogeneity, and/or edge currents that shunt the insulating bulk in transport measurements[91] and deserve further in-depth theoretical and experimental studies[85,92,93].

Moreover, inversion symmetry breaking engineered by the staggered sublattice potential can introduce Berry curvature hot spots near the PDP and SDC in monolayer graphene (left inset of Fig. 1d), allowing the realization of valley Hall effect[15,68,87]. Using a Hall bar geometry, the topological pure valley currents manifested as a nonlocal resistance by means of the inverse valley Hall effect (right inset of Fig. 1d) have been recently uncovered in commensurate graphene/*h*-BN heterostructures (Fig. 1d)[87]. Importantly, the amplitude of inversion symmetry breaking induced by staggered sublattice potential is highly tunable with interlayer coupling strength[88,94,95]. For example, decreasing the interlayer distance by using hydrostatic pressure produces an increase in the inversion symmetry breaking and bandgap in graphene/*h*-BN[94]. Apart from the *h*-BN substrate, a plethora of substrates, such as SiC[96], Ir(111)[97], Ni(111)[98] and Ag(111)[99] can produce staggered sublattice potential and break the inversion symmetry of monolayer graphene or other 2D materials, providing a firm basis for the development of new physics and applications[100].

**2.3 Interlayer twisting**

Due to the weak interlayer vdW interactions between the atomic planes, 2D layered materials can be stacked with tailor-made interlayer twist angle. The interlayer twist angle fundamentally determines the space group and thus presents an exotic degree of freedom for the modulation of crystal symmetry and interesting physical behaviors[50,101-103]. More specifically, for a given atomic layer, the operation of inversion symmetry can lead to only a purely translational displacement. Consequently, the inversion symmetry is explicitly broken for twisted bilayers constructed from centrosymmetric monolayers with a relative twist angle other than $\frac{2m\pi}{N}$, where $N$ denotes the $N$-fold rotational symmetry and $m$ is an integer[104]. For instance, for twisted bilayer graphene with $C_6$ rotational symmetry, the inversion centre between the top and bottom layers vanishes when the interlayer twist angle departs from 0° or multiples of 60°, resulting in the inversion symmetry breaking[105]. The manipulation of inversion symmetry breaking by interlayer twist angle leads to the observation of fascinating new phenomena, such as highly tunable second-order nonlinear optical responses (Fig. 1e)[102,105,106], ferroelectricity[107] and non-trivial valley polarization for both intralayer and interlayer excitons[101,108-110]. More significantly, the degree of inversion symmetry breaking and thus the corresponding physical properties can be continuously tuned by interlayer twist angle, ushering the age of twistronics and twistoptics for electronic, photonics and optoelectronic devices[50,56,104,111].

Recently, a breakthrough has been achieved in accurately controlling the interlayer twist angle, addressing the previous fabrication challenge that hinders the development of both basic scientific research and practical applications. Through the 'tear and stack' technique[112,113] or atomic force microscope (AFM) tip manipulation (Fig. 1f)[50,56,114], the interlayer twist angle can be controlled to 0.1° accuracy. Specifically, the AFM tip manipulation technique provides an *in situ* approach to dynamically control the interlayer twist angle in a single device. This can avoid extrinsic factors caused by device processing (e.g., interlayer coupling strength and cleanliness of the interface) and open up exciting opportunities to study the intrinsic twist angle-dependent properties and solve the nature of the existing puzzles. In addition, by combining the modified 'tear and stack' technique with highly oriented, wafer-scale, atomically thin 2D crystals, precise control of interlayer twist angle has been achieved in the centimeter scale[115], offering a firm basis for the development of practical twistronic applications.

**2.4 Crystal phase control**

For 2D materials that can exist in multiple polymorphs including both the centrosymmetric and inversion asymmetric structures, crystal phase control provides an effective route to engineer the underlying crystal symmetry. One notable example is transition metal dichalcogenides (TMDCs) which exhibit diverse crystal phases with distinct structures and symmetries, e.g., non-centrosymmetric 1H/$T_d$/3R phase and inversion symmetric 1T/1T′ phase (Fig. 1g)[110,116-118]. Through tuning the interactions between the lattices, spin, layers and strongly correlated electrons[117], it can trigger the structural phase transition between different polymorphs and thus provide on-demand control of the inversion symmetry breaking in TMDCs. For instance, phase transition between inversion asymmetric 1H or $T_d$ phase and centrosymmetric 1T or 1T′ phase has been demonstrated in atomically thin TMDCs under the driving force of electrostatic doping (Fig. 1h)[119,120], alkali metal intercalation[121], electron-beam irradiation[122], plasma bombardment[123], temperature[124-126], strain[127] and optical pump[128]. It is noteworthy that the local control of phase transition and symmetry breaking can lead to the realization of ohmic heterophase homojunction (Fig. 1i) for high-performance nanoelectronic devices with enhanced carrier mobility and on/off ratios[121,129].

### 2.5 Other methods

Since spin (valley) and velocity are axial and polar vectors respectively, the pure spin (valley) current possesses the odd symmetry under the operation of inversion symmetry[130]. Consequently, the pure spin (valley) current provides a low dissipation method to break the inversion symmetry. Recently, inversion symmetry breaking induced by pure spin current has been demonstrated in the bulk GaAs crystal[131]. Considering that 2D materials offer new playgrounds for generating pure spin (valley) current, successful control of inversion symmetry breaking with pure spin/valley currents in 2D systems might trigger a wave of fundamental studies. In addition, selectively patterned adsorption of atomic hydrogen onto one of the carbon sublattices or atom doping at one of the AB sublattices in monolayer graphene can break the equivalence of the two sublattices and inversion symmetry, potentially opening a sizable bandgap for electronic applications[100,132]. Similar strategies could be adopted by selective functionalization of other 2D materials and their heterostructures with different atoms and molecules (or even biological molecules).

Thus far, we have mainly focused on engineering the inversion symmetry breaking of 2D materials by controlling the crystal space group. The manipulation of magnetic point group (e.g., long-range spin ordering) also provides an exotic way to control the inversion symmetry breaking for magnetic 2D materials[5,133,134]. For example, for bilayer magnetic crystals with inversion-symmetric crystal space group, inversion symmetry is largely determined by the spin configurations. Inversion symmetry is broken and restored for layered antiferromagnetic order (upper panel of Fig. 1j) and ferromagnetic interlayer coupling (lower panel of Fig. 1j), respectively[6,10]. Recently, through controlling the layer-spin locking configurations by magnetic field[135], electric field[136] and electrostatic gating[137,138], inversion symmetry breaking in bilayer $CrI_3$ can be switched on or off. This facilitates continuous tuning of a wide variety of physical properties (such as second-order non-reciprocal optical effects[6], inelastic light scattering[10], bulk photovoltaic effect[139] and tunneling magnetoresistance[140-142]) for next-generation electronic and optoelectronic applications.

### 3. Engineering of $C_3$ rotational symmetry breaking in 2D materials

$C_3$ rotational symmetry breaking in 2D crystals can lead to anisotropic band structures, electronic properties and light-matter interactions, and thus sets a foundation for a wealth of intriguing physical phenomena and novel technological applications, such as in-plane anisotropy[143-145] and bulk photovoltaic effect[146,147], efficient energy harvesting devices[147,148] and polarization-sensitive photodetection[149]. Remarkably, recent studies have demonstrated that $C_3$ rotational symmetry breaking, together with inversion symmetry breaking, can lead to tilted Dirac cone and asymmetric distribution of Berry-curvature, ensuring nonvanishing Berry dipole, and thus various quantum geometrical phenomena, e.g., nonlinear Hall effect[11,77,150], valley magnetoelectricity[151] and valley orbital magnetization[152]. In this section, we discuss the typical engineering approaches to manipulate the $C_3$ rotational symmetry breaking for on-demand control of the properties in 2D materials.

### 3.1 Strain engineering

Strain engineering, a general strategy in semiconductor manufacturing to enhance device performance by introducing mechanical deformations in the crystal lattice, provides an excellent opportunity to control the $C_3$ rotational symmetry breaking in 2D materials. For example, when a uniaxial strain is applied onto a 2D crystal along certain crystal orientation (e.g., armchair or zigzag direction) by controllably bending the substrate (Fig. 2a), it can cause anisotropic lattice deformation (upper panel of Figs. 2b and 2c) and thus break the $C_3$ rotational symmetry. Note that the degree of $C_3$ rotational symmetry breaking can be continuously tuned over a large range due to the outstanding mechanical properties and stretchability of 2D materials. Significantly, $C_3$ rotational symmetry breaking engineered by uniaxial strain-field, together with intrinsic inversion symmetry breaking, enables the non-uniform distribution of Berry curvature and the emergence of in-plane Berry curvature dipole (***D***) around K/-K valleys in monolayer 2H-TMDCs (lower panel of Figs. 2b and 2c)[77,150]. Under the external in-plane electric field ***E***, the nonvanishing Berry curvature dipole ***D*** can serve as an effective internal magnetic field and give rise to the out-of-plane valley orbital magnetization $M \propto (D \cdot E)\hat{z}$ and nonlinear Hall effect $j(2\omega) \propto (D \cdot E)\hat{z} \times E$ [152]. Recently, highly tunable valley orbital magnetization has been uncovered in monolayer $MoS_2$ by controlling the $C_3$ rotational symmetry breaking and Berry curvature dipole with uniaxial strain (Figs. 2b and 2c)[151,152], providing an innovative route to generate and control magnetization in 2D materials. Apart from bending, uniaxial strain and $C_3$ rotational symmetry breaking can be controlled in various 2D materials for novel geometrical and topological quantum phenomena via stretching[77], triangular bubbles/holes[153] and corrugated structures[154,155]. It is worth noting that for homo- and hetero-structures with a small relative twist or lattice mismatch, the interplay between vdW interaction energy and elastic energy can induce an atomic lattice reconstruction, enabling an inhomogeneous in-plane strain profile to break the $C_3$ rotational symmetry (Fig. 2d)[156-159]. Apart from the control of $C_3$ rotational symmetry breaking, strain also offers a tempting strategy for engineering the electronic properties of 2D materials. For example, strain field with triangular symmetry can induce terribly large pseudo-magnetic fields over 300 Tesla and hence give rise to pseudo-magnetic quantum Hall effect in graphene[160,161], providing a firm basis for studying peculiar physics in the regime of extreme high magnetic fields.

**3.2 Effective dimensionality reduction**

$C_3$ rotational symmetry is explicitly broken in 1D systems due to the unique structural confinement between the axial and radial directions. Reducing the effective dimensionality of a 2D material into 1D hence can effectively introduce $C_3$ rotational symmetry breaking. For example, patterning the isotropic monolayer graphene[162] and $MoS_2$[163] into quasi-1D nanoribbons breaks the $C_3$ rotational symmetry and thus leads to the strong electrical and optical anisotropy. Recently, it was demonstrated that monolayer 2H-TMDCs can be scrolled into high-quality nanoscrolls or nanotubes (Fig. 2e)[147,164], unequivocally breaking the $C_3$ rotational symmetry. In conjunction with the broken inversion symmetry, TMDC nanotubes can exhibit non-vanishing shift currents and bulk photovoltaic effect[147,165]. Note that the efficiency of bulk photovoltaic effect is very high in TMDC nanotubes, and orders of magnitude larger than that in typical bulk materials[147], highlighting their potential in high-performance photovoltaics and other sustainable energy technologies.

**3.3 Phase transition**

In close analogy to the engineering of inversion symmetry breaking, structural phase transition also provides a way to control the $C_3$ rotational symmetry. For instance, the $C_3$ rotational symmetry breaking can be switched on or off by triggering the transition between different structural phases with and without $C_3$ rotational symmetry. Recent experiments show that via tuning the interlayer coupling and interlayer exchange interaction, structural transition from monoclinic stacking with antiferromagnetic interlayer coupling (Fig. 2g) to rhombohedral phase with ferromagnetic interlayer coupling (Fig. 2f) occurs in bilayer $CrI_3$, switching off the $C_3$ rotational symmetry breaking[166-168]. The inseparable connection between magnetic ground state and stacking order/$C_3$ rotational symmetry breaking offers a new insight toward the origin of exotic interlayer antiferromagnetic coupling in bilayer $CrI_3$[169,170].

**4. Engineering of time-reversal symmetry breaking in 2D materials**

Time-reversal symmetry breaking can lift the Kramers degeneracy for a wealth of physical phenomena, such as linear electrical Hall effect, magneto-circular dichroism and valley Zeeman effect. Through engineering time-reversal symmetry breaking, it is possible to realize on-demand control of these intriguing physical properties for various technological advances. In this section, we discuss the common engineering approaches to control the time-reversal symmetry breaking in 2D materials.

**4.1 Magnetic field**

In contrast to the polar feature of electric field, magnetic field is axial in nature and can break the time-reversal symmetry and Kramers doublet. The broken time-reversal symmetry under out-of-plane magnetic fields can break the odd characteristics of Berry curvature $\Omega(k)$ with respect to $k$ ($\Omega(k) = -\Omega(-k)$) and lead to non-vanishing transverse Hall conductivity given by the integral of $\Omega(k)$ over the entire Brillouin-zone (BZ) $\sigma_H = \frac{e^2}{\hbar} \int_{BZ} \frac{d^2k}{(2\pi)^2} f_k \Omega(k)$, where $f_k$ is the equilibrium

Fermi-Dirac distribution[72]. Consequently, linear electrical Hall effect can be observed in various 2D materials under vertical magnetic fields (Fig. 3a)[1,13,18,85,171]. Notably, for a closed manifold, the path integral of $\mathit{\Omega}(k)$ is quantized and the transverse Hall conductivity $\sigma_H$ is the Chern number in the units of $\frac{e^2}{h}$, known as the quantum Hall effect (QHE)[72]. The QHE was first demonstrated in monolayer graphene (Fig. 3b)[13,171] and subsequently observed in a wide range of atomically thin crystals under an external magnetic field, such as bilayer/trilayer graphene[172,173], black phosphorus[174] and InSe[175].

More significantly, due to the valley-dependent magnetic moment in inversion-asymmetric 2D Dirac materials (e.g., gapped graphene and monolayer 2H-TMDCs), time-reversal symmetry breaking induced by magnetic fields opens up an exciting opportunity to lift the valley degeneracy through Zeeman-like interaction, known as the valley Zeeman effect (Fig. 3c)[44,45]. The breaking of valley degeneracy sets a new research topic for the development of valley-polarized photocurrents and valleytronic logics[176]. Usually, the valley Zeeman splitting is described by the non-interacting independent-particle picture with a dominant contribution of intracellular magnetic moment from intra-atomic $d$ orbital and is about 0.23 meV T$^{-1}$ (corresponding to a Landé $g$-factor around 4)[44-46]. Such a small valley magnetic response strongly impedes the magnetic manipulation of valley pseudospin degrees of freedom. Recent advances demonstrate that the engineering of Coulomb interactions with electron doping can drive the electron-hole symmetry breaking and first-order magnetic phase transition[177,178], giving rise to strongly enhanced valley Zeeman splitting with a net intercellular contribution from phase winding of the Bloch wavepacket[179,180]. Meanwhile, nearly two orders of magnitude enhanced valley magnetic response has been recently observed in WSe$_2$/WS$_2$ moiré superlattices by taking advantage of the antiferromagnetic interactions between localized excitons and holes around half-filling of the first hole moiré superlattice band[181], opening exciting possibilities for valleytronic applications.

**4.2 Optical control**

Coherent light-matter interaction provides an exotic way to optically control the symmetry breaking. Two salient examples associated with time-reversal symmetry breaking are valley-selective optical stark effect and Bloch-Siegert shift in 2D materials with inversion-asymmetric hexagonal lattices (Fig. 3d)[182]. An equilibrium system driven coherently by off-resonance light could produce a series of photon-dressed states (Floquet states) with energy spacing in units of the photon energy. In the first order approximation, there are two pairs of Floquet states, one is between the two original states and the other is outside the original states. These two types of Floquet states can hybridize with the equilibrium Bloch states, leading to state repulsion and transition energy shift. The former and latter cases are known as the optical Stark effect (left panel of Fig. 3d) and Bloch-Siegert effect (right panel of Fig. 3d), respectively[182]. Due to valley-contrasting magnetic quantum numbers and matching condition of angular momentum, the optical Stark effect and Bloch-Siegert effect in monolayer 2H-TMDCs are valley-exclusive and obey opposite selection rules. For instance, excited with $\sigma^+$ ($\sigma^-$) radiation, the optical Stark effect exclusively occurs at K (-K) valley, while the Bloch-Siegert shift is at -K (K) valley. Therefore, by pumping with a circularly polarized light, valley-exclusive optical Stark effect and Bloch-Siegert shift would break the time-reversal symmetry and lift the valley degeneracy. Indeed, an energy splitting of >10 meV between the K and -K excitons has been demonstrated by both optical Stark effect[183,184] and Bloch-Siegert shift[182]. Such a splitting is an order of magnitude larger than the Zeeman shift commonly accessible by static magnetic fields[44,45], highlighting the great potential in optical manipulation of time-reversal symmetry breaking and valley degree of freedom. Moreover, the manipulation of valley pseudospin has been demonstrated at femtosecond timescale by valley-selective optical Stark effect in monolayer 2H-TMDCs[185], underscoring the possibility of ultrafast all-optical valleytronics.

**4.3 Magnetic proximity effect**

Magnetic proximity effect can transform a given nonmagnetic material through its adjacent magnetic substrate to become magnetic, consequently breaking the time-reversal symmetry[186,187]. Recently, time-reversal symmetry breaking induced by proximity magnetic exchange field has been demonstrated in atomically thin 2D crystals (e.g., graphene, $MoSe_2$ and $WSe_2$) via the detection of anomalous Hall effect[188], Zeeman spin Hall effect[189] and spontaneous valley Zeeman splitting[187,190-192]. Benefiting from the atomic thickness and sharp interfaces between 2D materials and their adjacent magnetic substrates, 2D materials can harbor a substantial magnetic exchange field, in contrast to their bulk counterparts where the size dwarfs the characteristic lengths of proximity coupling[186,187]. Indeed, valley Zeeman splitting with one order of magnitude enhancement has been realized in monolayer $WSe_2$ by utilizing the substantial proximity magnetic effects from ferromagnetic EuS[190] and CrI$_3$[191] substrates, offering potential applications in valleytronics. Moreover, owing to the strong magnetic proximity effect, nonmagnetic 2D materials with remarkable spin-dependent physical properties can be exploited as a sensitive magnetic sensor to probe the subtle magnetic domain structures for the adjacent magnetic regions. For instance, by utilizing the spin-valley properties of monolayer WSe$_2$ as a magnetic probe (Fig. 3e), layered antiferromagnetic domain structures in bilayer/trilayer CrI$_3$, for which the conventional magnetometry techniques are difficult to detect, have been successfully mapped out (Fig. 3f)[187].

**4.4 Magnetic atom doping**

Magnetic atom doping can induce magnetism and break the time-reversal symmetry in 2D materials. For example, through doping magnetic transition metal elements (such as Cr and Fe, Fig. 3g) into 2D topological insulators (e.g., Bi$_2$Te$_3$, Bi$_2$Se$_3$, and Sb$_2$Te$_3$), spontaneous magnetic moments and time-reversal symmetry breaking can be achieved[193,194]. The broken time-reversal symmetry, together with band inversion transition induced by strong spin-orbit coupling, can give rise to topologically nontrivial electronic structure characterized by a finite Chern number, leading to the realization of quantum anomalous Hall effect in 2D magnetic topological insulators[193]. Recently, via fine-tuning the Fermi level around the magnetically induced energy gap, quantum anomalous Hall effect has been experimentally uncovered in 2D Cr-doped (Bi$_{0.1}$Sb$_{0.9}$)$_{1.85}$Te$_3$, enabling the developing of low-power-consumption electronics[37]. In addition to 2D topological insulators, magnetic atom doping can also control the time-reversal symmetry breaking of other 2D materials, such as graphene[100], MoS$_2$[195] and SnS$_2$[196].

**4.5 Other methods**

Since charge current can cause the carrier scattering and dissipation, it can introduce time-reversal symmetry breaking in 2D conductors[151]. For instance, when a charge current is applied to a strained MoS$_2$ monolayer, time-reversal symmetry breaking induced by current, in conjunction with the broken inversion and $C_3$ rotational symmetries, can lead to robust valley magnetoelectricity[151]. In addition, a wide range of chemical strategies, such as defects, surface functionalization and decoration, can break time-reversal symmetry in 2D materials[197]. For instance, hydrogen atoms adsorbed on graphene can lead to the prohibition of double occupation of $\pi$-state by two electrons with different spins, breaking the time-reversal symmetry to induce a magnetic moment[198].

## 5. Engineering spontaneous gauge symmetry breaking in 2D materials

Spontaneous gauge symmetry breaking provides the necessary ingredient for Bose-Einstein condensation and superconductivity[199]. In this section, we present the common approaches to engineer the spontaneous gauge symmetry breaking in 2D materials.

### 5.1 Carrier doping

In 2D materials, the formation of degenerate carriers can enable strong Coulomb interactions and break spontaneous gauge symmetry, offering an effective way to access the exotic superconducting states[200]. Recently, by using ionic-liquid-gating to achieve high carrier densities inside, spontaneous gauge symmetry breaking and superconducting dome have been demonstrated in atomically thin $MoS_2$ and $WS_2$ (Fig. 4a)[31,200-203]. Moreover, inversion symmetry breaking in monolayer $MoS_2$ and $WS_2$ can give rise to the Zeeman-type spin splitting with a large valley-contrasting effective out-of-plane magnetic field of hundreds of teslas (Fig. 4b)[31]. Consequently, the spin configurations of electrons in Cooper pairs are locked to out-of-plane directions. This can strongly quench the Pauli paramagnetic effect and enhance the in-plane pair-breaking field[29,31]. Indeed, unconventional Ising superconductivity with in-plane upper critical field far beyond the Pauli paramagnetic limit has been recently demonstrated in 2D 2H-TMDCs[29,31,202,204,205].

### 5.2 Flat-band engineering

Spontaneous symmetry breaking can typically occur in many-body systems. Creating flat bands with high density of states in the momentum space can largely reduce the electron lattice hopping kinetic energy and drive the system into strong electron-electron interactions, enabling spontaneous gauge symmetry breaking and a rich phase diagram of correlation physics[181,206-208]. In vdW homo- and hetero-structures formed by stacking two 2D crystals, rotational misalignment or lattice mismatch introduces an in-plane periodic moiré potential (Fig. 4c), which provides an exotic route to achieve flat minibands (Fig. 4d) and break spontaneous gauge symmetry[206,209]. Recently, spontaneous U(l) gauge symmetry breaking and superconducting domes flanking the partial moiré band filling Mott state (Figs. 4e and 4f) have been observed in magic-angle twisted bilayer/double bilayer graphene[28,209-212], ABC-stacked trilayer graphene/$h$-BN moiré superlattices[207] and twisted bilayer $WSe_2$[213]. Such *in-situ* gate-tunable 2D systems with relatively simple chemical composition open up exciting opportunities to solve the puzzle of strong-correlation Hubbard physics and understand the nature of high-transition-temperature superconductivity.

To achieve the long-sought mechanism of unconventional superconductivity, it is also important to unravel other spontaneous symmetry breaking in magic-angle systems. Recently, spontaneous time-reversal symmetry breaking was revealed by significant ferromagnetic hysteresis and giant anomalous Hall effect in magneto-transport behavior[211,214-216]. Interestingly, the Hall resistivity is well-quantized to von Klitzing constant, demonstrating the exotic quantum anomalous Hall effect and non-trivial Chern number for topological moiré miniband[215,216]. Through scanning tunneling spectroscopy[217-219], it was found that the spatial charge modulation within AA region exhibits a quadrupole geometry for the partial flat band filling, consisting of four quadrants with alternating positive and negative charge. Such local charge ordering would be aligned to form global stripe charge order, clearly breaking the $C_3$ rotational symmetry of the moiré superlattice[217]. At the same time, some results show that $C_2$ and $C_2T$ (defined as a 180 degree rotation combined with time reversal) symmetries are also spontaneously broken in magic-angle twisted graphene[211,220,221]. Nevertheless, this is just the beginning and further studies are required to demystify all the broken symmetry in magic-angle twisted graphene, which could provide essential clues to the nature of the correlated electronic states and unconventional superconductivity.

**6: Perspectives and conclusions**

The scientific and technological advances in 2D materials and their mixed-dimensional structures have introduced a wide variety of methods to engineer their underlying symmetry breaking. Control of symmetry breaking could lead to exotic physics in highly symmetrical 2D materials without intrinsic symmetry breaking, greatly expanding the candidate materials for basic scientific research and advanced applications. Moreover, the manipulation of symmetry breaking can continuously tune the macroscopic electronic properties and lead to fascinating physical phenomena, paving the way toward understanding the puzzles in basic physics and developing new research for emerging technological innovations.

**6.1 Novel structures**

The properties of 2D materials are usually severely limited by their high symmetry. Importantly, on-demand control of symmetry breaking can be achieved by specially-designing various exotic structures (e.g., vdW heterostructures, stretched structures, twisted structures, mixed-dimensional structures and localized doping). As a consequence, fascinating new phenomena and exciting applications that are not possible for the intrinsic 2D materials without symmetry breaking can be introduced. For example, the most studied graphene represents an unique platform for topological valley currents as it possesses ultrahigh electron mobility and two inequivalent valley pseudospin states at the corners of the BZ[2]. However, the presence of inversion symmetry in graphene makes it impossible to control the valley pseudospin by optical helicity or electromagnetic field. Through engineering inversion symmetry breaking via staggered sublattice potential[87] or out-of-plane electric field[57], robust topological valley transport up to room temperature and over long distances has been realized in monolayer/bilayer graphene, demonstrating a robustness suitable for valleytronic applications. Moreover, there is no doubt that the unique integration possibilities offered by mixing 2D materials and 2D/non-2D materials or various novel micro- and nano-structures continuously present substantial future growth potential in engineering symmetry breaking for both fundamental understanding and applications[52].

**6.2 New physics**

It is well-known that symmetry breaking is usually accompanied by the emergence of fascinating phenomena. Through engineering symmetry breaking, different broken symmetries can be integrated together, leading to synergistic effects and creating exotic opportunities to underpin new physics. This is particularly important to understand the nature of the existing puzzles, explore emerging quantum phases, introduce novel quantum degrees of freedom, initiate new research fields and stimulate technological advances in novel photonics, electronics, spintronics, valleytronics and twistronics. For example, via engineering various spontaneous symmetry breaking with electron-electron correlations in twisted graphene moiré superlattices, semimetallic graphene could be turned into a series of quantum states (e.g., band insulator, Mott-like correlated insulator, quantum anomalous Hall insulator and superconductor), potentially offering new insights into the well-known conundrums in strongly correlated physics, such as high-temperature unconventional superconductors, magnetic insulators, quantum spin liquids and frustrated antiferromagnets.

Remarkably, one of the most exciting directions in the near future is to engineer the symmetry breaking of magic-angle Moiré superlattices through different external degrees of freedom (e.g., electric field, doping and pressure), which will further initiate exotic quantum states and transform the landscape of condensed-matter physics. To date, engineering of symmetry breaking by precise control of twist angle in Moiré superlattices has enabled a rich quantum phase diagram of strong correlated behaviours (e.g., exciton Bose-Einstein condensation[30,222], unconventional superconductivity[28,211], quantized anomalous Hall states[215,216], nematic/stripe phases[217,223,224], chiral spin density wave[225] and Wigner crystal states[226-229]) just in the span of less than three years, and will undoubtedly bring more surprises and exotic insight into various quantum phase transitions in the future[230]. Although hugely promising progresses have been achieved, grand challenges still remain. One of the obstacles is the ubiquity of sizable twist-angle inhomogeneity from the fabrication of devices[230,231]. In fact, due to the twist-angle disorder (where a small fluctuation can result in sizable inhomogeneity in physical properties), there have not yet been reports of two identical results. This is a serious hindrance towards the establishment of a unified understanding and elucidation of the underlying physical mechanism. In addition, for practical applications, large scale Moiré superlattices with uniform twist-angle are required. The current 'tear-and-stack' based fabrication method[112], however, usually realizes only micron-sized sample. This requires the development of new techniques, especially the controllable and reproducible bottom-up growth, to address this issue.

**6.3 Emerging technological innovation**

Engineering of symmetry breaking in 2D materials and their mixed-dimensional structures have already defined a bright vision for a wide range of new and emerging applications, such as intriguing optoelectronic/photonic devices[232], exotic valleytronic/twistronic devices[4,15,56], novel energy devices[147] and ultralow-power-consumption electronics[37,74]. Meanwhile, we highlight that the engineering of symmetry breaking can also be applied to integrated devices/systems beyond 2D materials, such as 1D materials[233], optical cavities[234-236], sonic crystals[237,238], photonic lattices[239,240], cold-atom systems[241,242], introducing new horizons for applications. For instance, topological valley transport and Weyl quasiparticles have been recently demonstrated in phononic and photonic crystals by engineering symmetry breaking[237-239], potentially for macroscopic-scale topological applications. The insight of symmetry breaking engineering into massive interdisciplinary fields would give rise to the interweaving and stimulation of different research directions and may open exciting opportunities for not only new physics and their applications, but also enabling transformative advances beyond physics, e.g., quantum computation, asymmetric catalysis, efficient energy harvesting, communications and medicine.

**Acknowledgements**

The authors gratefully acknowledge the financial support by Academy of Finland (Grant Nos. 314810, 333982, 336144, 336818 and 333099), Academy of Finland Flagship Programme (PREIN), the EU H2020-MSCA-RISE-872049 (IPN-Bio), the European Union's Horizon 2020 research and innovation program (Grant No. 820423, S2QUIP) and ERC (Grant No. 834742). C.N.L acknowledges the support of DOE BES DE-SC0020187, NSF DMR 1807928 and 1922076.

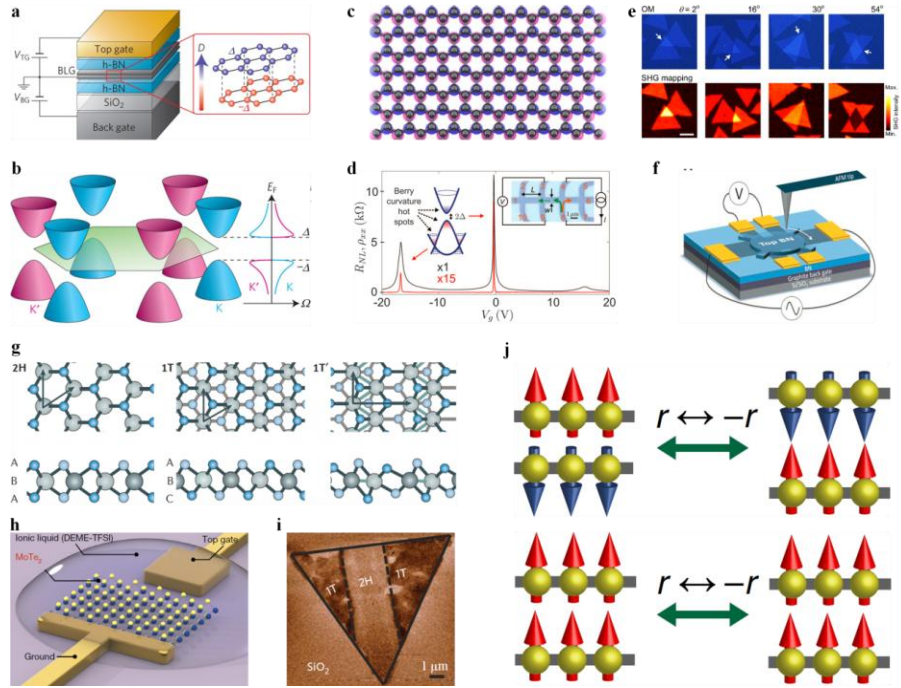

**Fig. 1 | Engineering the inversion symmetry breaking. a,** Left: schematic diagram of dual-gated bilayer graphene device. Right: conceptual pictures for interlayer potential energy difference and inversion symmetry breaking induced by a vertical electric field. **b**, Band structure (left) and Berry curvature (right) of bilayer graphene with broken inversion symmetry. **c,** Schematic diagram of graphene/$h$-BN moiré superlattices with staggered sublattice potential. **d,** Nonlocal resistance (red curve) and longitudinal resistance (black curve) in aligned graphene/$h$-BN superlattices. Left (right) inset is the band structure with Berry curvature hot spots (nonlocal measurement geometry). **e,** Interlayer twisting approach for tuning second harmonic generation in twisted $MoS_2$ bilayers (upper: optical microscopy images; lower: second harmonic generation intensity mapping). **f,** Schematic cartoon of the AFM-tip manipulation of interlayer twist angle. **g,** Top (upper) and side (lower) view of the crystal structure for different polymorphs of monolayer TMDCs. **h,** Schematic of crystal phase control with ionic liquid gating in monolayer $MoTe_2$. **i,** Electrostatic force microscopy phase image of a phase-engineered $MoS_2$ homojunction. **j,** Schematics of the spin configurations dependent inversion symmetry. Inversion symmetry is broken and restored for layered antiferromagnetic (upper) and ferromagnetic (lower) coupling. Panels **a** and **b** adapted from REF.[64]. Panel **d** adapted from REF.[87]. Panel **e** adapted from REF.[102]. Panel **f** adapted from REF.[50]. Panel **g** adapted from REF.[116]. Panel **h** adapted from REF.[120]. Panel **i** adapted from REF.[121]. Panel **j** adapted from REF.[10].

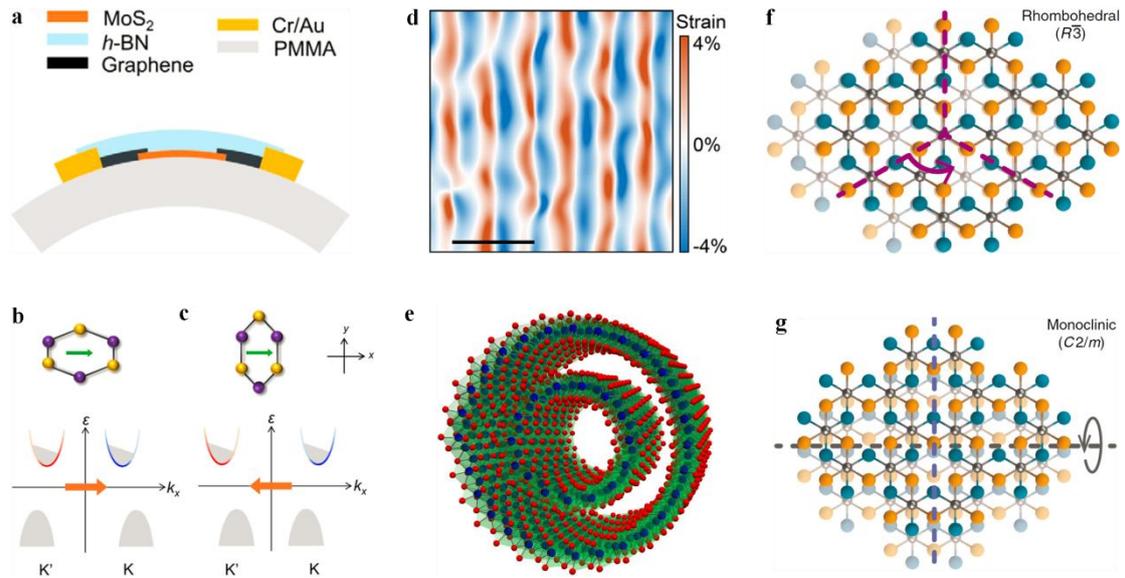

**Fig. 2 | Engineering the $C_3$ rotational symmetry breaking. a,** Schematic of the strain-engineered device structure with a bending polymethyl methacrylate (PMMA) substrate. **b,c,** Upper: top view of the MoS$_2$ crystal structure under strain along zigzag (**b**) and armchair (**c**) directions. Lower: the corresponding modified band structure and Berry curvature distribution of upper panels. Red (blue) colors represent positive (negative) Berry curvature. Orange arrows denote the Berry curvature dipole. **d,** Example of inhomogeneous in-plane strain profile and hence $C_3$ rotational symmetry breaking in a heterostructure enabled by the interplay between vdW interaction energy and elastic energy. Scale bar: 10 nm. **e,** Illustrations of a quasi-1D WS$_2$ multiwall nanotube. **f,g,** Atomic structure of bilayer CrI$_3$ with rhombohedral (**f**) and monoclinic (**g**) stacking. The rhombohedral structure possesses an out-of-plane $C_3$ axis. Panels **a-c** adapted from REF.[152]. Panel **d** adapted from REF.[159]. Panel **e** adapted from REF.[147]. Panels **f** and **g** adapted from REF.[6].

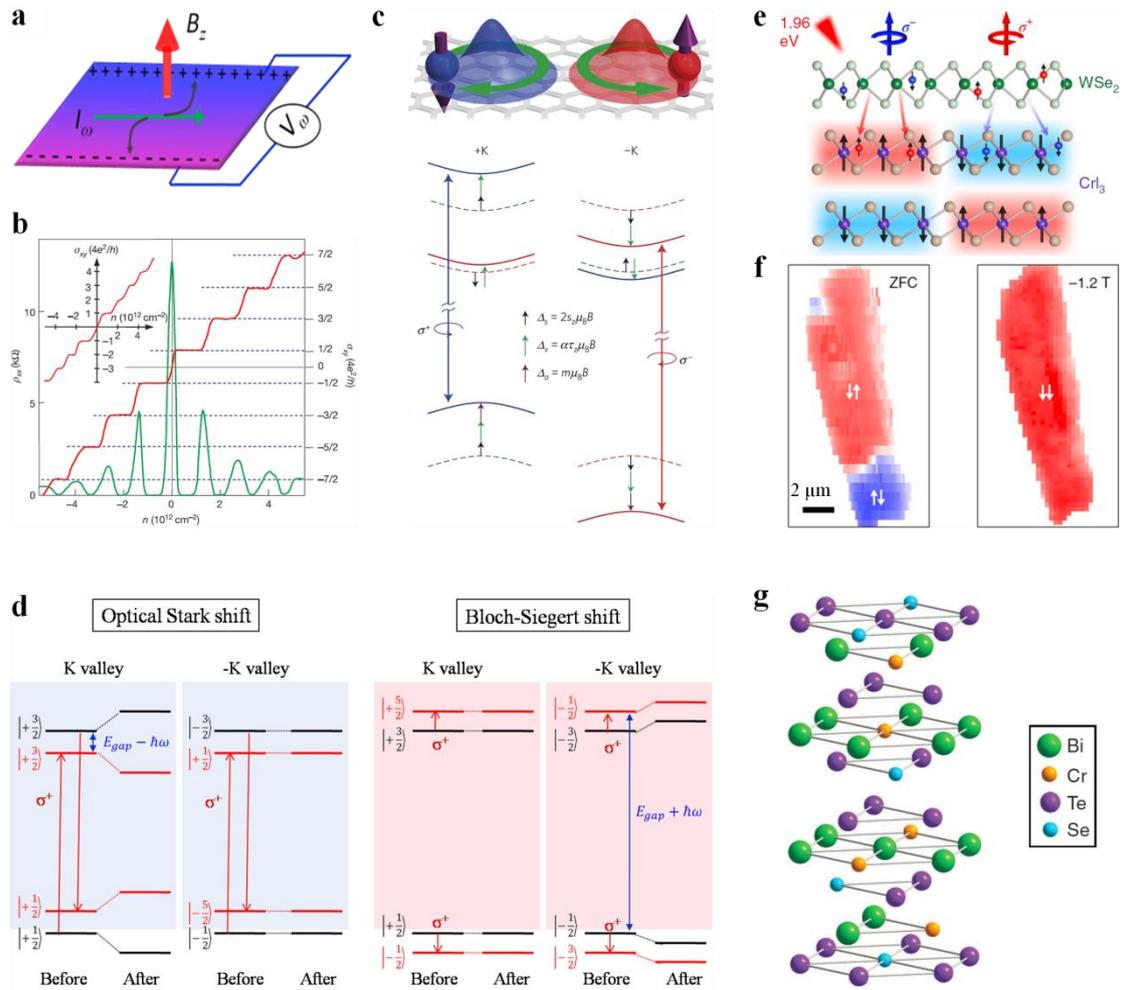

**Fig. 3 | Engineering the time-reversal symmetry breaking. a,** Schematic diagram for Hall effect introduced by magnetic field. **b,** QHE for massless Dirac fermions of monolayer graphene. **c,** Upper: cartoon depicting the self-rotation of the wavepacket (valley magnetic moments). Lower: energy level diagram for magnetic field induced valley Zeeman effect with three contributions from spin (black), valley (green) and atomic orbital (purple), respectively. **d,** Energy diagram for valley dependent optical Stark shift (left) and Bloch-Siegert shift (right) in monolayer 2H-TMDCs. **e,** Monolayer $WSe_2$ with exotic spin-valley properties as a sensitive magnetic sensor to probe the layered antiferromagnetic domain structures of adjacent bilayer $CrI_3$. **f,** Spatial map of the spontaneously circular polarization of monlayer $WSe_2$ at zero magnetic field (left) and -1.2 T (right). **g,** Schematic crystal structure of Cr-doped $Bi_2(Se_xTe_{1-x})_3$. Panel **b** adapted from REF.[13]. Panel **c** adapted from REF.[45]. Panel **d** adapted from REF.[182]. Panels **e** and **f** adapted from REF.[187]. Panel **g** adapted from REF.[194].

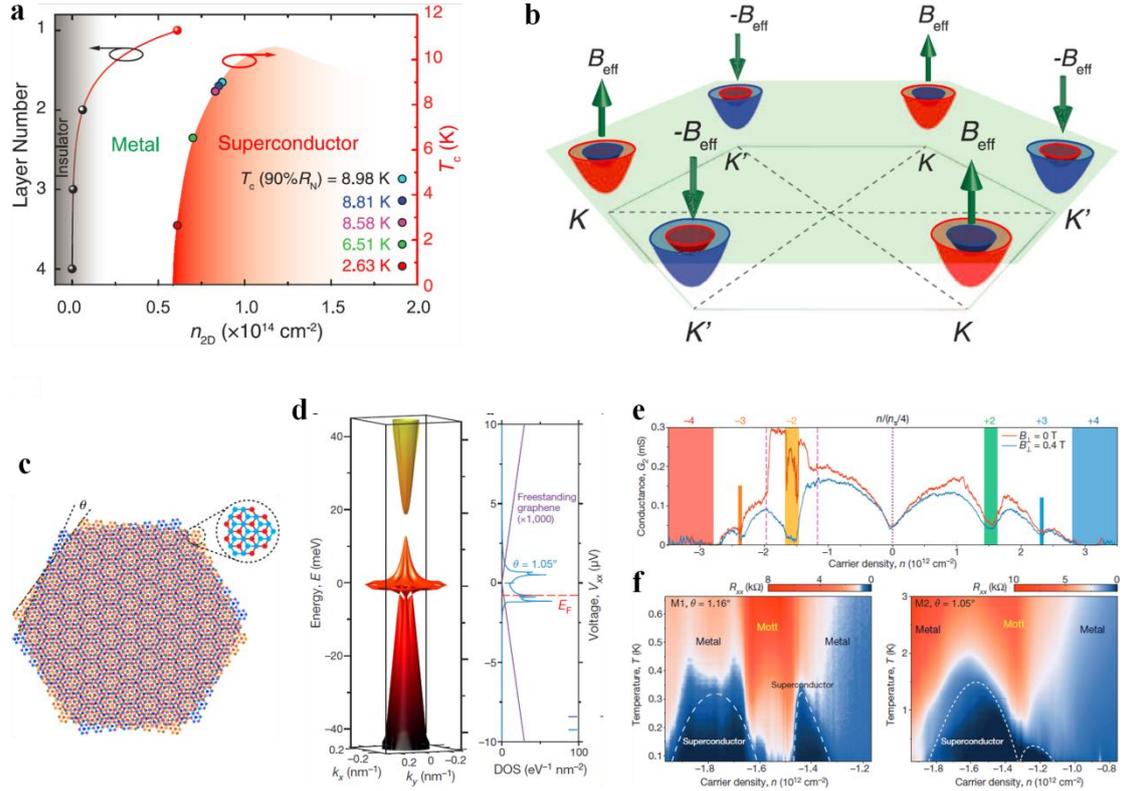

**Fig. 4 | Engineering the spontaneous gauge symmetry breaking. a,** Phase diagram of superconductivity as a function of carrier doping density. **b,** Schematic image of the valley-dependent spin polarization. Electrons in K and -K valleys experience opposite effective magnetic fields $+B_{eff}$ and $-B_{eff}$, respectively. **c,** Schematic of the moiré superlattices of magic-angle double bilayer graphene. **d,** Periodic moiré potential engineered flat bands near charge neutrality with width less than 15 meV. **e,** Two-probe longitudinal conductance versus carrier density at zero magnetic field (red) and 0.4 T (blue) for device with twist angles of 1.16°. **f,** Colour plot of longitudinal resistance as a function of carrier density and temperature for devices with twist angles of 1.16° (left) and 1.05° (right), showing metal, correlated Mott state and superconducting phase. Panels **a** and **b** adapted from REF.[31]. Panel **c** adapted from REF.[209]; Panels **d-f** adapted from REF.[28].